\documentclass[10pt]{article}

\usepackage{amsmath,amsthm,verbatim,amssymb,amsfonts,amscd, graphicx}
\usepackage{graphics}
\theoremstyle{plain}
\usepackage{amsmath} 
\usepackage{braket}

\pdfoutput=1

\title{Efficient learning algorithm for quantum perceptron unitary weights}

\author{
 Kok-Leong Seow\\Department of Electrical Engineering and Computer Science\\Wichita State University\\kxseow1@wichita.edu
  \and Elizabeth C. Behrman\\Department of Mathematics, Statistics, and Physics\\Wichita State University\\elizabeth.behrman@wichita.edu
  \and James E. Steck\\Department of Aerospace Engineering\\Wichita State University\\ james.steck@wichita.edu
}
\date{}

\begin{document}

\maketitle


\begin{abstract}
	For the past two decades, researchers have attempted to create a Quantum Neural Network (QNN) by combining the merits of quantum computing and neural computing. In order to exploit the advantages of the two prolific fields, the QNN must meet the non-trivial task of integrating the unitary dynamics of quantum computing and the dissipative dynamics of neural computing. At the core of quantum computing and neural computing lies the qubit and perceptron, respectively. We see that past implementations of the quantum perceptron model have failed to fuse the two elegantly. This was due to a slow learning rule and a disregard for the unitary requirement. In this paper, we present a quantum perceptron that can compute functions uncomputable by the classical perceptron while analytically solving for parameters and preserving the unitary and dissipative requirements. 

\end{abstract}

{\bf Keywords} Quantum computing, Artificial neural networks, Unitary, Quantum neural networks, Singular value decomposition


\section{Introduction}
The idea of quantum mechanics being computationally more powerful than a classical Turing machine was first proposed by Feynman in \cite{feynman1986quantum, feynman1982simulating}. Thus far, there have been only two algorithms that utilize the power of quantum computing: Shor's factorization \cite{shor1994algorithms} and Grover's database search \cite{grover1996fast}. As quantum computing technology evolves, it seems natural to combine the properties of artificial neural networks with quantum computing. The synergy of quantum and neural computing will grant us advantages unattainable by classical artificial networks.

As stated in \cite{ezhov2000quantum}, quantum neural networks have advantages in the following areas: 1) exponential memory capacity \cite{ventura2000quantum}, 2) higher performance for lower number of hidden neurons \cite{cutting1999would}, 3) faster learning \cite{menneer1995quantum}, 4) elimination of catastrophic forgetting due to the absence of pattern interference \cite{menneer1995quantum}, 5) single layer network solution of linear inseparable problems \cite{menneer1995quantum}, 6) absence of wires \cite{behrman1999spatial}, 7) processing speed \cite{behrman1999spatial}, 8) small scale \cite{behrman1999spatial}, and 9) higher stability and reliability \cite{cutting1999would}. While Kak introduced the concept of applying quantum mechanics to artifical neural networks \cite{kak1995quantum}, the first in depth examination of QNNs were provided in Menneer's PhD thesis \cite{menneer1999quantum}. Since then, a myriad of attempts to combine the strengths of quantum computing and neural computing have been presented, and are reviewed in \cite{schuld2014quest}. 

In the review, Shuld  \textit{et al.} classified all QNN models into 4 different approaches: 1) interpreting the step-function as measurement in order to combine the non-linearity of neural networks and linearity of quantum computing, 2) using quantum circuits that attempt to model the dynamics of neural networks, 3) describing the basic component of NNs, the perceptron, with a quantum formalism, and 4) modeling NNs with interacting quantum dots consisting of four atoms sharing two electrons. In this paper, we focus on the third approach of creating a quantum equivalent for a perceptron, with the goal to construct the building block for a more complex quantum neural network. 

The first attempt to develop a quantum perceptron to overcome the limitations imposed by classical perceptrons was in 2001 by Altaisky \cite{altaisky2001quantum}. Altaisky implemented his quantum perceptron through the use of optical modes, optical beam splitters, and phase shifters. Altaisky directly translated the perceptron model and replaced it with $\Ket{y} = \hat{F}\sum_{j=1}^{N}{\hat{w}_j\Ket{x_j}}$ along with the quantum equivalent of the learning rule, 
\begin{equation}
\hat{w}_j(t+1) = \hat{w}_j(t) + \eta(\Ket{d} - \Ket{y(t)})\Bra{x_j}.
\label{eq:rule}
\end{equation}
Following Altaisky was Fei \textit{et al.} \cite{fei2003study}, who implemented a new version of the iterative rule that could compute the XOR-function using one neuron, a feat not capable by classical perceptrons. Fei's activation operator and weights were unitary, but provided no way to derive the operators/weights which seemed specific to each problem. Next, Zhou \textit{et al.} \cite{zhou2007quantum} also computed the XOR-function with one neuron, but required many iterations. Its learning rule implied a possible non unitary gate, contradicting the requirements of quantum computing. Slightly different from the previous models, Siomau introduced an autonomous quantum perceptron based on calculating a set of positive valued operators and value measurements (POVM) \cite{siomau2014quantum}. However, his perceptron could not replicate basic quantum gates such as quantum-Not gate and Hadamard gate, which are staples of quantum computing. Finally, Sagheer and Zidan \cite{sagheer2013autonomous} implemented their Autonomous Quantum Perceptron Neural Network by calculating the set of activation operators using the form: $\left( \begin{smallmatrix} \cos{\theta_j} & -\sin{\theta_j} \\ \sin{\phi_j} & \cos{\phi_j} \end{smallmatrix} \right)$. Similar to the aforementioned models, it also has an iterative method to learn the weights, as well as the issue of the learned weights/operators being non unitary. This severe violation creates a need for our proposed quantum perceptron that removes the iterative learning and analytically solves for weights/operators while maintaining the unitary requirements.

This paper is organized as follows: Section 2 describes the background mathematics. Topics included are the basics of quantum computing, neural networks, and singular value decomposition (SVD). Section 3 presents the the quantum perceptron and learning algorithm. We illustrate the perceptron's robustness by first applying the basic learning rule to the NOT gate and Hadamard gate to show that the newfound weights/operators took several iterations and are not unitary, which thus violates a strict requirement. To combat this issue, we present a closed-form solution to find the proper weights by using Moore-Penrose pseudoinverse and a tensor product. We illustrate that this also generalizes to two qubits and is capable of computing the controlled NOT function. We observe that in the case of computing our nonlinear controlled swap gate, that the unitary isn't preserved. We fix the issue through the use of SVD, and illustrate it's significance by computing the XOR function. We conclude in section 4.

 
\section{Background}
\subsection{Quantum Computing}
The basic elements of quantum computing are qubits, complex unit vectors in a 2-dimensional Hilbert space $\mathcal{H}_2$. Labeled $\ket{0}$ and $\ket{1}$, the two quantum states express one bit of information: $\ket{0}$ corresponds to the bit 0 of classical computers, and $\ket{1}$ to the bit 1. More formally, given the basis $\{\ket{0}, \ket{1}\}$, a basic qubit is given in Dirac notation and described by the wavefunction 
\begin{equation}
\label{eq:equationQuantum}
\begin{split}
&\ket{\psi} = \alpha\ket{0} + \beta\ket{1}, \\
&  \ \ \ \ \ \ \ \alpha,\beta \in \mathbb{C}
\end{split}
\end{equation}
where $\alpha$ and $\beta$ are complex numbers called probability amplitudes that satisfy $|\alpha|^2 + |\beta|^2 = 1$. The state $\psi$ of a general quantum system is in a linear superposition of basis states $\ket{0}$ and $\ket{1}$ with $|\alpha|^2$ probability of collapsing into state $\ket{0}$ and $|\beta|^2$ probability of collapsing into state $\ket{1}$. The qubits may also be written in vector notation, $\ket{0} = \left( \begin{smallmatrix} 1 \\ 0 \end{smallmatrix} \right)$ and $\ket{1} = \left( \begin{smallmatrix} 0 \\ 1 \end{smallmatrix} \right)$

\subsection{Neural Networks}

Neurons are symbolized by variables $x,y=\{-1,1\}$ where '1' indicates that the neuron is firing and 
'-1' indicates that it is resting. The activation mechanism of a neuron $y$ due to the input of $N$ other neurons forms the core of neural computing. This setup creates the simplest neural network possible: the perceptron. The perceptron is a computational model of a single neuron and is shown in figure \ref{fig:perceptron}. Given $n$ input nodes called neurons with values $x_k = \{-1,1\}, k = 1 \dots n$ that feed signals into a single output neuron $y$. Each input neuron is connected to the output neuron with a certain strength denoted by weight parameter $w_k \in [-1,1)$. Neuron $y$ is activated in an 'integrate-and-fire' mechanism, meaning that the post-synaptic signals are simply added up and compared with the specific threshold $\theta_y$ of neuron $y$. The input-output relation is governed by the activation function
\begin{equation}
\label{eq:equationPerceptron}
y = \left\{ \begin{array}{rcl}
1, & \mbox{if} & \sum_{k=1}^{m}{w_{k}x_k} \geq \theta_y \\ 
-1 & \mbox{else}  \\
\end{array}\right.
\end{equation} 
The perceptron learning rule is governed by the equation
\begin{equation}
\label{eq:equationTwo}
w_k(t+1) = w_k(t)+\eta (d-y)x
\end{equation}
Worth mentioning, the classical perceptron is unable to calculate the non-separable XOR function. 
\subsection{Singular Value Decomposition}
Any rectangular matrix $A$ can be decomposed into three matrices defined as
\begin{equation}
A = U\Sigma V^T
\end{equation}
where $U \in \mathbb{R}^{N\times N}$ is a unitary matrix, i.e., $U^TU = UU^T = I$ (identity matrix), $\Sigma \in \mathbb{R}^{N\times N}$ is diagonal matrix of singular values $\sigma_1 \geq \sigma_2 \geq \ldots \geq \sigma_r > 0$ , and $V \in \mathbb{R}^{M\times M}$ is a unitary matrix $V^TV = VV^T = I$ (identity matrix).

\section{Quantum Neural Network}
To describe the efficiency of our proposed quantum neuron and algorithm, we first present the quantum learning rule given in Eq. \ref{eq:rule} to see where issues arise. We then counter the iterative method by analytically solving for weights by using Moore-Penrose pseudoinverse, a similar technique used by Extreme Learning Machine (ELM) \cite{huang2004extreme} in the classical domain. Finally, to preserve the quantum properties of being unitary, we decompose our non-unitary matrix $\hat{w}$ into three unitary matrices using our version of SVD.
\subsection{Moore-Penrose Pseudoinverse}		
Given a quantum perceptron of the form $\Ket{y_{output}} = \hat{F}\sum_{j=1}^{N}{\hat{w}_j\Ket{x_j}}$, we can set $\hat{F}=I$ and obtain $\Ket{y_{output}} =\sum_{j=1}^{N}{\hat{w}_j\Ket{x_j}}$. We can then find the Moore-Penrose pseudoinverse of $x$, and analytically calculate $\hat{w}_j$ through the equation
\begin{equation}
\hat{w_j} = \ket{y_j} \otimes \ket{x_j^\dagger}
\label{eq:moore}
\end{equation} 
where we take the tensor product of $\ket{y_j}$ and $\ket{x_j^\dagger}$.
We first apply equation \ref{eq:rule} to the NOT gate and Hadamard gate to show the possibility of non-unitary weights and slow convergence.
\subsubsection{The NOT gate}
Given two training examples
\begin{equation}
\begin{split}
& \ket{x_1} = \ket{0}, \ \ket{y_1} = \ket{1} \\
& \ket{x_2} = \ket{1}, \ \ket{y_2} = \ket{0}
\end{split}
\notag
\end{equation}
utilizing the learning rule, we set $\hat{F}=I$ and randomize the initial weights to $w_0 =\left( \begin{smallmatrix} 1 & 0 \\ 0 & -1 \end{smallmatrix} \right)$, $\eta = 0.1$, and target output $\ket{y}=\ket{0}$. Applying equation \ref{eq:rule}, we get
\begin{equation}
\begin{split}
\hat{w}(t+1) & = \left( \begin{matrix} 1 & 0 \\ 0 & -1 \end{matrix} \right) + 0.1(\ket{0}+\ket{1})\bra{1} \\
& = \left( \begin{matrix} 1 & 0 \\ 0 & -1 \end{matrix} \right) + 0.1(\ket{0}\bra{1}+\ket{1}\bra{1}) \\
& = \left( \begin{matrix} 1 & 0 \\ 0 & -1 \end{matrix} \right) + \left( \begin{matrix} 0 & 0.1 \\ 0 & 0.1 \end{matrix} \right)\\
& = \left( \begin{matrix} 1 & 0.1 \\ 0 & -0.9 \end{matrix} \right)
\end{split}
\notag
\end{equation}
after one iteration. We see that after 25 iterations we obtain
\begin{equation}
\begin{split}
\hat{w} = \left( \begin{matrix} 1 & 0.9282102 \\ 0 & -0.0717898 \end{matrix} \right) \ \ \ \ \ \ \ \ \ \ \ \ \ \ \ \ \ \ \ \  \\
\mbox{and} \ket{y_{output}} = \left( \begin{matrix} 1 & 0.9282102 \\ 0 & -0.0717898 \end{matrix} \right) \left( \begin{matrix} 0 \\ 1 \end{matrix} \right) = \left( \begin{matrix} 0.9282102 \\ -0.0717898 \end{matrix} \right).
\end{split}
\notag
\end{equation}
It is easy to see that the calculated weights are not unitary, violating an important property of quantum computing. Instead, we calculate the Moore-Penrose pseudoinverse of the input and determine weights analytically. We apply equation Eq. \ref{eq:moore} to inputs $\ket{x_1}$ and $\ket{x_2}$. \begin{equation}
\begin{split}
\ket{x_1^\dagger} = \left( \begin{matrix} 1 & 0 \end{matrix} \right) \\
\hat{w_1}=\ket{y_1} \otimes \ket{x_1^\dagger} = \left( \begin{matrix} 0 & 0 \\ 1 & 0 \end{matrix} \right) \\
\ket{x_2^\dagger} = \left( \begin{matrix} 0 & 1 \end{matrix} \right) \\
\hat{w_2}=\ket{y_2} \otimes \ket{x_2^\dagger} = \left( \begin{matrix} 0 & 0 \\ 0 & 1 \end{matrix} \right) \\
\hat{w} = \sum_{j=1}^{N}{\hat{w_j}} = \left( \begin{matrix} 0 & 1 \\ 0 & 0 \end{matrix} \right) + \left( \begin{matrix} 0 & 0 \\ 1 & 0 \end{matrix} \right) = \left( \begin{matrix} 0 & 1 \\ 1 & 0 \end{matrix} \right)
\end{split}
\notag
\end{equation}.
Using our new weights, we compute our output.
\begin{equation}
\begin{split}
\Ket{y_{output}} = \sum_{j=1}^{N}{\hat{w}\Ket{x_1}} = \left( \begin{matrix} 0 & 1 \\ 1 & 0 \end{matrix} \right) \left( \begin{matrix} 1 \\ 0 \end{matrix} \right) = \left( \begin{matrix} 0 \\ 1 \end{matrix} \right) = \ket{1} \\
\Ket{y_{output}} = \sum_{j=1}^{N}{\hat{w}\Ket{x_2}} = \left( \begin{matrix} 0 & 1 \\ 1 & 0 \end{matrix} \right) \left( \begin{matrix} 0 \\ 1 \end{matrix} \right) = \left( \begin{matrix} 1 \\ 0 \end{matrix} \right) = \ket{0}
\notag
\end{split}
\end{equation}
It is easy to see that our newly learned weights are indeed unitary, and after only one iteration.

\subsubsection{The Hadamard gate}
Given two training examples 
\begin{equation}
\begin{split}
& \ket{x_1} = \ket{0}, \ \ket{y_1} = \frac{1}{\sqrt{2}}\ket{0} + \frac{1}{\sqrt{2}}\ket{1} \\
& \ket{x_2} = \ket{1}, \ \ket{y_2} = \frac{1}{\sqrt{2}}\ket{0} - \frac{1}{\sqrt{2}}\ket{1} 
\end{split}
\notag
\end{equation}
by utilizing the learning rule, we set $\hat{F}=I$ and randomize the initial weights to $w_0 =\left( \begin{smallmatrix} 1 & 0 \\ 0 & -1 \end{smallmatrix} \right)$, $\eta = 0.1$, and target output $\ket{y}=\frac{1}{\sqrt{2}}\ket{0} + \frac{1}{\sqrt{2}}\ket{1} = \left( \begin{smallmatrix} \frac{1}{\sqrt{2}} \\ \frac{1}{\sqrt{2}} \end{smallmatrix} \right)$. We see that after 25 iterations we obtain
\begin{equation}
\begin{split}
\hat{w} = \left( \begin{matrix} 0.72813353 & 0 \\ 0.65634373 & -1 \end{matrix} \right) \ \ \ \ \ \ \ \ \ \ \ \ \ \ \ \ \ \ \ \  \\
\mbox{and} \ket{y_{output}} = \left( \begin{matrix} 0.72813353 & 0 \\ 0.65634373 & -1 \end{matrix} \right) \left( \begin{matrix} 1 \\ 0 \end{matrix} \right) = \left( \begin{matrix} 0.72813353 \\ 0.65634373 \end{matrix} \right).
\end{split}
\notag
\end{equation}
A quick observation shows us that these weights also violate the unitary condition. We again calculate the Moore-Penrose pseudoinverse of the input and determine weights analytically. We apply Eq. \ref{eq:moore} to inputs $\ket{x_1}$ and $\ket{x_2}$.
\begin{equation}
\begin{split}
\ket{x_1^\dagger} = \left( \begin{matrix} 1 & 0 \end{matrix} \right) \\
\hat{w_1}=\ket{y_{1}} \otimes \ket{x_1^\dagger} = \left( \begin{matrix} \frac{1}{\sqrt{2}} & 0 \\ \frac{1}{\sqrt{2}} & 0 \end{matrix} \right) \\
\ket{x_2^\dagger} = \left( \begin{matrix} 0 & 1 \end{matrix} \right) \\
\hat{w_2}=\ket{y_{2}} \otimes \ket{x_2^\dagger} = \left( \begin{matrix} 0 & \frac{1}{\sqrt{2}} \\ 0 & -\frac{1}{\sqrt{2}} \end{matrix} \right) \\
\hat{w} = \sum_{j=1}^{N}{\hat{w_j}} = \left( \begin{matrix} \frac{1}{\sqrt{2}} & 0 \\ \frac{1}{\sqrt{2}} & 0 \end{matrix} \right) + \left( \begin{matrix} 0 & \frac{1}{\sqrt{2}} \\ 0 & -\frac{1}{\sqrt{2}} \end{matrix} \right) = \left( \begin{matrix} \frac{1}{\sqrt{2}} & \frac{1}{\sqrt{2}} \\ \frac{1}{\sqrt{2}} & -\frac{1}{\sqrt{2}} \end{matrix} \right)
\end{split}
\notag
\end{equation}
Using our computed weights, we determine our output
\begin{equation}
\begin{split}
\Ket{y_{output}} = \hat{F}\sum_{j=1}^{N}{\hat{w_j}\Ket{x_1}} = \left( \begin{matrix} \frac{1}{\sqrt{2}} & \frac{1}{\sqrt{2}} \\ \frac{1}{\sqrt{2}} & -\frac{1}{\sqrt{2}} \end{matrix} \right) \left( \begin{matrix} 1 \\ 0 \end{matrix} \right) = \left( \begin{matrix} \frac{1}{\sqrt{2}} \\ \frac{1}{\sqrt{2}} \end{matrix} \right) = \frac{1}{\sqrt{2}}\ket{0} + \frac{1}{\sqrt{2}}\ket{1} \\
\Ket{y_{output}} = \hat{F}\sum_{j=1}^{N}{\hat{w_j}\Ket{x_2}} =\left( \begin{matrix} \frac{1}{\sqrt{2}} & \frac{1}{\sqrt{2}} \\ \frac{1}{\sqrt{2}} & -\frac{1}{\sqrt{2}} \end{matrix} \right) \left( \begin{matrix} 0 \\ 1 \end{matrix} \right) = \left( \begin{matrix} \frac{1}{\sqrt{2}} \\ -\frac{1}{\sqrt{2}} \end{matrix} \right) = \frac{1}{\sqrt{2}}\ket{0} - \frac{1}{\sqrt{2}}\ket{1}
\notag
\end{split}
\end{equation}
It's easily verifiable that our determined weights are indeed unitary, and after only one iteration.
\subsubsection{The Controlled NOT gate}
We show that the Moore-Penrose pseudoinverse is also applicable to 2-qubit systems.
Given four training examples 
\begin{equation}
\begin{split}
& \ket{x_1} = \ket{00}, \ \ket{y_1} = \ket{00} \\
& \ket{x_2} = \ket{01}, \ \ket{y_2} = \ket{01}  \\
& \ket{x_3} = \ket{10}, \ \ket{y_3} = \ket{11} \\
& \ket{x_4} = \ket{11}, \ \ket{y_4} = \ket{10},  \\
\end{split}
\notag
\end{equation}
after setting $\hat{F} = I$, we then calculate the Moore-Penrose psuedoinverse.

\begin{equation}
\begin{split}
\ket{x_1^\dagger} = \left( \begin{matrix} 1 & 0 & 0 & 0 \end{matrix} \right) \\ 
\ket{x_2^\dagger} = \left( \begin{matrix} 0 & 1 & 0 & 0 \end{matrix} \right) \\ 
\ket{x_3^\dagger} = \left( \begin{matrix} 0 & 0 & 1 & 0 \end{matrix} \right) \\ 
\ket{x_4^\dagger} = \left( \begin{matrix} 0 & 0 & 0 & 1 \end{matrix} \right) \\
\end{split}
\end{equation}
We then take the tensor product of the pseudoinverse and input to create its corresponding weight. Summing up of all the individual weights, we acquire  the final weight.
\begin{equation}
\begin{split}
\hat{w_1}=\ket{y_1} \otimes \ket{x_1^\dagger} = \left( \begin{matrix} 1 & 0 & 0 & 0 \\ 0 & 0 & 0 & 0 \\ 0 & 0 & 0 & 0 \\ 0 & 0 & 0 & 0 \end{matrix} \right) \
\hat{w_2}=\ket{y_2} \otimes \ket{x_2^\dagger} = \left( \begin{matrix} 0 & 0 & 0 & 0 \\ 0 & 1 & 0 & 0 \\ 0 & 0 & 0 & 0 \\ 0 & 0 & 0 & 0 \end{matrix} \right) \\
\hat{w_3}=\ket{y_3} \otimes \ket{x_3^\dagger} = \left( \begin{matrix} 0 & 0 & 0 & 0 \\ 0 & 0 & 0 & 0 \\ 0 & 0 & 0 & 0 \\ 0 & 0 & 0 & 1 \end{matrix} \right) \
\hat{w_4}=\ket{y_4} \otimes \ket{x_4^\dagger} = \left( \begin{matrix} 0 & 0 & 0 & 0 \\ 0 & 0 & 0 & 0 \\ 0 & 0 & 0 & 1 \\ 0 & 0 & 0 & 0 \end{matrix} \right) \\
\hat{w} = \sum_{j=1}^{N}{\hat{w_j}} = \left( \begin{matrix} 1 & 0 & 0 & 0 \\ 0 & 1 & 0 & 0 \\ 0 & 0 & 0 & 1 \\ 0 & 0 & 1 & 0 \end{matrix} \right) 
\end{split}
\notag
\end{equation}
By observation, it is easy to see that our algorithm generalizes to the 2-qubit system and are indeed unitary.

\subsubsection{The Controlled Swap gate}
We now test the dissipative features of our qubit by attempting to calculate the nonlinear version of the controlled swap gate. If qubit one is HIGH then swap qubits 2 and 3. Our eight states are shown below:
\begin{equation}
\begin{split}
& \ket{x_1} = \ket{000}, \ \ket{y_1} = \ket{00}, \ \ket{x_2} = \ket{001}, \ \ket{y_2} = \ket{01} \\
& \ket{x_3} = \ket{010}, \ \ket{y_3} = \ket{10},  \ \ket{x_4} = \ket{011}, \ \ket{y_4} = \ket{11} \\
& \ket{x_5} = \ket{100}, \ \ket{y_5} = \ket{00}, \ \ket{x_6} = \ket{101}, \ \ket{y_6} = \ket{10} \\
& \ket{x_7} = \ket{110}, \ \ket{y_7} = \ket{01},  \ \ket{x_8} = \ket{111}, \ \ket{y_8} = \ket{11}. \\
\end{split}
\notag
\end{equation}
After calculating the weights by searching for the Moore-Penrose pseudoinverse and tensoring, we sum up all of the individual weights.
\begin{equation}
\hat{w} = \sum_{j=1}^{N}{\hat{w_j}} = \left( \begin{matrix} 1 & 0 & 0 & 0 & 1 & 0 & 0 & 0 \\ 0 & 1 & 0 & 0 & 0 & 1 & 0 & 0\\ 0 & 0 & 1 & 0 & 0 & 1 & 0 & 0 \\ 0 & 0 & 0 & 1 & 0 & 0 & 0 & 1 \end{matrix} \right)
\notag
\end{equation}
We see that in the nonlinear case, the weight matrix isn't unitary. In order to combat this, we introduce Singular Value Decomposition (SVD). SVD returns two unitary matrices and a diagonal matrix of singular values. We observe that there isn't a loss of generality by removing the diagonal matrix and replacing it with a unitary matrix of ones in the diagonals and zeros everywhere else. By setting
\begin{equation}
\hat{F}\Sigma_{new}\hat{w_{new}} = U\Sigma V^T 
\end{equation}
we can thus decompose the $\hat{w}$ into three matrices. To improve accuracy we can also introduce a measurement function $M(x)$ such that 
\begin{equation}
\label{eq:measurement}
M = \left\{ \begin{array}{rcl}
1, & \mbox{if} & x > 0.5 \\ 
0 & \mbox{else}  \\
\end{array}\right.
\end{equation} 
Applying SVD to $\hat{w}$ we obtain
\begin{equation}
\small
\begin{split}
\hat{F} = \left( \begin{matrix} 1 & 0 & 0 & 0 \\ 0 & 1 & 0 & 0\\ 0 & 0 & 1 & 0 \\ 0 & 0 & 0 & 1 \end{matrix} \right), \ \Sigma_{new} = \left( \begin{matrix} 1 & 0 & 0 & 0 & 0 & 0 & 0 & 0 \\ 0 & 1 & 0 & 0 & 0 & 0 & 0 & 0 \\ 0 & 0 & 1 & 0 & 0 & 0 & 0 & 0 \\ 0 & 0 & 0 & 1 & 0 & 0 & 0 & 0 \end{matrix} \right), \\ \hat{w_{new}} = \left( \begin{matrix} \frac{1}{\sqrt{2}} & 0 & 0 & 0 & \frac{1}{\sqrt{2}} & 0 & 0 & 0 \\ 0 & \frac{1}{\sqrt{2}} & 0 & 0 & 0 & 0 & \frac{1}{\sqrt{2}} & 0 \\ 0 & 0 & \frac{1}{\sqrt{2}} & 0 & 0 & \frac{1}{\sqrt{2}} & 0 & 0 \\ 0 & 0 & 0 & \frac{1}{\sqrt{2}} & 0 & 0 & 0 & \frac{1}{\sqrt{2}} \\ -\frac{1}{\sqrt{2}} & 0 & 0 & 0 & \frac{1}{\sqrt{2}} & 0 & 0 & 0 \\ 0 & 0 & -\frac{1}{\sqrt{2}} & 0 & 0 & \frac{1}{\sqrt{2}} & 0 & 0 \\ 0 & -\frac{1}{\sqrt{2}} & 0 & 0 & 0 & 0 & \frac{1}{\sqrt{2}} & 0 \\ 0 & 0 & 0 & -\frac{1}{\sqrt{2}} & 0 & 0 & 0 & \frac{1}{\sqrt{2}} \end{matrix} \right)
\end{split}
\notag
\end{equation}
Calculating 
\begin{equation}
M(\hat{F}\Sigma_{new}\hat{w_{new}}\ket{x}),
\label{eq:newRule}
\end{equation}
we consequently get the proper outcome in a closed form solution while maintaining unitary weights.
\section{Proposed Algorithm}
In this section we summarize the full algorithm and demonstrate its prowess by computing the XOR function.
\\ Given $N$ data sets:
\\ 1. Set $\hat{F}=I$ (Identity matrix). 
\\ 2. Calculate the Moore-Penrose pseudoinverse of each input.
\\ 3. Take the tensor product of the pseudoinverse of each input and its respective target to obtain the corresponding weight using Eq. \ref{eq:moore}.
\\ 4. Sum up all of the weights to create weight matrix $\hat{w}$
\\ 5. If $\hat{w}$ is not unitary, move to step 6, else move to step 8.
\\ 6. Use SVD to decompose the $\hat{w}$ into three matrices $\hat{F}\Sigma\hat{w}_{new}$
\\ 7. Pull out the diagonal matrix of singular values and replace it with a unitary matrix $\Sigma_{new}$ with ones in the diagonals and zeroes elsewhere. 
\\ 8. Use Eq. \ref{eq:newRule} to calculate $\ket{y_{output}}$.

\subsubsection{The XOR function}
The XOR function is nonlinear and uncomputable by a single neuron in the classical perceptron. Given a set of four different inputs and targets
\begin{equation}
\begin{split}
& \ket{x_1} = \ket{00}, \ \ket{y_1} = \ket{0} \\
& \ket{x_2} = \ket{01}, \ \ket{y_2} = \ket{1}, \\ 
& \ket{x_3} = \ket{10}, \ \ket{y_3} = \ket{1}, \\ 
& \ket{x_4} = \ket{11}, \ \ket{y_4} = \ket{0}, \\ 
\end{split}
\notag
\end{equation}
we can place the inputs in a superposition governed by the wave function $\ket\psi$.
\begin{equation}
\ket{\psi} = \frac{1}{\sqrt{4}}\ket{00} + \frac{1}{\sqrt{4}}\ket{01} + \frac{1}{\sqrt{4}}\ket{10} + \frac{1}{\sqrt{4}}\ket{11} 
\notag
\end{equation}
After garnering the pseudoinverse of each state and tensoring with its respective target, we sum up all of the individual weights to obtain out final weight.
\begin{equation}
\hat{w} = \sum_{j=1}^{N}{\hat{w}_j} = \left( \begin{matrix} \frac{1}{\sqrt{4}} & 0 & 0 & \frac{1}{\sqrt{4}} \\ 0 & \frac{1}{\sqrt{4}} & \frac{1}{\sqrt{4}} & 0 \end{matrix} \right)
\notag
\end{equation}
Decomposing $\hat{w}$ into three unitary matrices, we now acquire
\begin{equation}
\begin{split}
\hat{F} = \left( \begin{matrix} 1 & 0 \\ 0 & 1 \end{matrix} \right), \ \Sigma_{new} = \left( \begin{matrix} 1 & 0 & 0 & 0 \\ 0 & 1 & 0 & 0 \end{matrix} \right), 
\\ \hat{w_{new}} = \left( \begin{matrix} \frac{1}{\sqrt{2}} & 0 & 0 & \frac{1}{\sqrt{2}} \\ 0 & \frac{1}{\sqrt{2}} & \frac{1}{\sqrt{2}} & 0 \\ 0  & -\frac{1}{\sqrt{2}} & \frac{1}{\sqrt{2}} & 0 \\ -\frac{1}{\sqrt{2}} & 0 & 0 & \frac{1}{\sqrt{2}} \end{matrix} \right).
\end{split}
\notag
\end{equation}
Calculating $M(\hat{F}\Sigma_{new}\hat{w_{new}}\ket{x})$ returns the proper output.


\section{Conclusion}
To harness the computational power of quantum computing and artificial neural networks, fundamental rules that grant them their intrinsic properties must not be infringed. In addition, QNNs should be robust enough to not be overly sensitive to the choice of operators as well as the ability to analytically solve for its parameters, similar to its classical counterpart. The work of the quantum perceptron model for the past two decades have failed to demonstrate these imperative qualities. In our proposed quantum perceptron and algorithm, we expel the iterative learning rule by analytically calculating parameters while maintaining unitary requirements. We also demonstrate its nonlinear capabilities by computing the XOR function, a feat impossible for the classical perceptron.

\newpage
\nocite{*}
\bibliography{QNNbib}

\begin{thebibliography}{10}

\bibitem{feynman1986quantum}
R.~P. Feynman, ``Quantum mechanical computers,'' {\em Foundations of physics},
  vol.~16, no.~6, pp.~507--531, 1986.

\bibitem{feynman1982simulating}
R.~P. Feynman, ``Simulating physics with computers,'' {\em International
  journal of theoretical physics}, vol.~21, no.~6/7, pp.~467--488, 1982.

\bibitem{shor1994algorithms}
P.~W. Shor, ``Algorithms for quantum computation: Discrete logarithms and
  factoring,'' in {\em Foundations of Computer Science, 1994 Proceedings., 35th
  Annual Symposium on}, pp.~124--134, IEEE, 1994.

\bibitem{grover1996fast}
L.~K. Grover, ``A fast quantum mechanical algorithm for database search,'' in
  {\em Proceedings of the twenty-eighth annual ACM symposium on Theory of
  computing}, pp.~212--219, ACM, 1996.

\bibitem{ezhov2000quantum}
A.~A. Ezhov and D.~Ventura, ``Quantum neural networks,'' in {\em Future
  directions for intelligent systems and information sciences}, pp.~213--235,
  Springer, 2000.

\bibitem{ventura2000quantum}
D.~Ventura and T.~Martinez, ``Quantum associative memory,'' {\em Information
  Sciences}, vol.~124, no.~1, pp.~273--296, 2000.

\bibitem{cutting1999would}
D.~Cutting, ``Would quantum neural networks be subject to the decidability
  constraints of the church-turing thesis,'' {\em Neural Network World},
  vol.~9, pp.~163--168, 1999.

\bibitem{menneer1995quantum}
T.~Menneer and A.~Narayanan, ``Quantum-inspired neural networks,'' {\em
  Department of Computer Science, University of Exeter, Exeter, United Kingdom,
  Technical Report}, vol.~329, p.~1995, 1995.

\bibitem{behrman1999spatial}
E.~C. Behrman, J.~E. Steck, and S.~R. Skinner, ``A spatial quantum neural
  computer,'' in {\em Neural Networks, 1999. IJCNN'99. International Joint
  Conference on}, vol.~2, pp.~874--877, IEEE, 1999.

\bibitem{kak1995quantum}
S.~Kak, ``On quantum neural computing,'' {\em Information Sciences}, vol.~83,
  no.~3, pp.~143--160, 1995.

\bibitem{menneer1999quantum}
T.~S.~I. Menneer, {\em Quantum artificial neural networks.}
\newblock PhD thesis, University of Exeter, 1999.

\bibitem{schuld2014quest}
M.~Schuld, I.~Sinayskiy, and F.~Petruccione, ``The quest for a quantum neural
  network,'' {\em Quantum Information Processing}, vol.~13, no.~11,
  pp.~2567--2586, 2014.

\bibitem{altaisky2001quantum}
M.~Altaisky, ``Quantum neural network,'' {\em arXiv preprint quant-ph/0107012},
  2001.

\bibitem{fei2003study}
L.~Fei and Z.~Baoyu, ``A study of quantum neural networks,'' in {\em Neural
  Networks and Signal Processing, 2003. Proceedings of the 2003 International
  Conference on}, vol.~1, pp.~539--542, IEEE, 2003.

\bibitem{zhou2007quantum}
R.~Zhou and Q.~Ding, ``Quantum mp neural network,'' {\em International Journal
  of Theoretical Physics}, vol.~46, no.~12, pp.~3209--3215, 2007.

\bibitem{siomau2014quantum}
M.~Siomau, ``A quantum model for autonomous learning automata,'' {\em Quantum
  information processing}, vol.~13, no.~5, pp.~1211--1221, 2014.

\bibitem{sagheer2013autonomous}
A.~Sagheer and M.~Zidan, ``Autonomous quantum perceptron neural network,'' {\em
  arXiv preprint arXiv:1312.4149}, 2013.

\bibitem{huang2004extreme}
G.-B. Huang, Q.-Y. Zhu, and C.-K. Siew, ``Extreme learning machine: a new
  learning scheme of feedforward neural networks,'' in {\em Neural Networks,
  2004. Proceedings. 2004 IEEE International Joint Conference on}, vol.~2,
  pp.~985--990, IEEE, 2004.

\bibitem{behrman2000simulations}
E.~C. Behrman, L.~Nash, J.~E. Steck, V.~Chandrashekar, and S.~R. Skinner,
  ``Simulations of quantum neural networks,'' {\em Information Sciences},
  vol.~128, no.~3, pp.~257--269, 2000.

\bibitem{ventura1998artificial}
D.~Ventura and T.~Martinez, ``An artificial neuron with quantum mechanical
  properties,'' in {\em Artificial Neural Nets and Genetic Algorithms},
  pp.~482--485, Springer, 1998.

\bibitem{zhou2006quantum}
R.~Zhou, L.~Qin, and N.~Jiang, ``Quantum perceptron network,'' in {\em
  Artificial Neural Networks--ICANN 2006}, pp.~651--657, Springer, 2006.

\bibitem{narayanan2000quantum}
A.~Narayanan and T.~Menneer, ``Quantum artificial neural network architectures
  and components,'' {\em Information Sciences}, vol.~128, no.~3, pp.~231--255,
  2000.

\bibitem{panella2011neural}
M.~Panella and G.~Martinelli, ``Neural networks with quantum architecture and
  quantum learning,'' {\em International Journal of Circuit Theory and
  Applications}, vol.~39, no.~1, pp.~61--77, 2011.

\bibitem{rosenblatt1958perceptron}
F.~Rosenblatt, ``The perceptron: a probabilistic model for information storage
  and organization in the brain.,'' {\em Psychological review}, vol.~65, no.~6,
  p.~386, 1958.

\end{thebibliography}
\bibliographystyle{ieeetr}
\end{document}